\documentstyle[12pt]{article}

\newcommand{\dslash}{\partial \hspace{-.5em}/\hspace{.2em}}

\newcommand{\Aslash}{A \hspace{-.5em}/\hspace{.2em}}

\newcommand{\beq}{\begin{eqnarray}}
\newcommand{\enq}{\end{eqnarray}}
\newcommand{\beqn}{\begin{eqnarray*}}
\newcommand{\enqn}{\end{eqnarray*}}
\newcommand{\rar}{\rightarrow}

\newcommand{\non}{\nonumber}

\def\eg {{\it e.g., }}

\def\bitem{\begin{itemize}}
\def\eitem{\end{itemize}}
\def\thefootnote{\fnsymbol{footnote}}

\begin{document}
\begin{titlepage}
\begin{flushright}
SUNY-NTG-94-37
\end{flushright}
\vskip 0.6in
\centerline{\large \bf        
Heavy Mesons in a Random Instanton Gas
\footnote{Supported by the Department of Energy
under Grant No.\, DE-FG02-88ER40388. }}
\vskip 0.6in
\centerline{S. Chernyshev, M. A. Nowak and I. Zahed}
\vskip 0.4in
\centerline{\it Department of Physics}
\centerline{\it SUNY at Stony Brook}
\centerline{\it Stony Brook, New York 11794}
\vskip 1.25in

\centerline{\bf ABSTRACT}
\vskip .5cm

We analyze  the correlation function of a meson with one heavy
and one light quark in inverse powers of the
heavy quark mass $m_Q$ using a succession of
Foldy-Wouthuysen-type transformations
prior to radiative corrections. We evaluate the correlator to order
$m_Q^0$ in a random and dilute gas of instantons, using the planar
approximation. We show, in leading order in the density, that the heavy
quark mass is shifted to the order $m_Q^0$ and that the induced interaction
between the heavy and light quarks is attractive. We also find it to be an
order of magnitude smaller than the 't Hooft interaction between two light
quarks. The shift in the heavy quark mass is related to the perimeter law of
large Wilson loops. The relevance of these results for general hadronic
correlators with heavy quarks is discussed.

\vfill
\begin{flushleft}
SUNY-NTG-94-37\\
July 1994
\end{flushleft}
\vfill
\end{titlepage}

\renewcommand{\thefootnote}{\#\arabic{footnote}}
\setlength{\baselineskip}{27pt}
\setcounter{footnote}{0}
\setcounter{equation}{0}

The instanton model for the QCD ground state \cite{shur}
offers an interesting framework
for the discussion of soft physics from first principles. While instantons
do not confine color in pure Yang-Mills theory, they provide a simple
mechanism for the spontaneous breaking of chiral symmetry in QCD. Extensive
calculations using instantons both analytically and numerically suggest that
the instantons in the vacuum form a dilute gas \cite{dia}.
This may be understood
by noting that the instanton-antiinstanton interaction is not dipole like at
large distances but screened by the fermions over distances on the order of
$1/2$ fm. A dilute system of instantons behaves as a free gas.
As a result, the singlet $\eta^\prime$ is heavy and the topological
susceptibility is small.

Hadrons with a heavy quark exhibit a new type of symmetry:  invariance under
spin flip of the heavy quark. This symmetry has been used in combination with
chiral symmetry (Heavy Quark Effective Theory) \cite{hqet}
to understand the structure
and decay properties of heavy-light systems. The main observation is that if
the mass of one quark is taken to infinity and factorized, the  physics of
the remaining system is soft and constrained by chiral dynamics.
The  analysis of the soft physics in heavy-light systems can be
assessed quantitatively using QCD lattice simulations, or qualitatively
using QCD inspired models.

In this letter, we  investigate the effects of a dilute gas of instantons
on mesons composed of one heavy and one light quark.
We show that the heavy
meson correlator may be systematically investigated in inverse powers of the
heavy quark mass using standard Foldy-Wouthuysen transformations \cite{fw}
and before the inclusion of radiative corrections.
We analyze the heavy meson correlator in the planar approximation and for
a dilute gas of instantons. To leading order in the density, the instantons
shift the heavy quark mass as first noted by Diakonov,
Petrov and Pobylitsa \cite{diapob},
and induce an attractive scalar-color interaction between
the heavy and light quarks that is an order of magnitude weaker than the
induced 't Hooft interaction between two light quarks \cite{hoo}. The
shift in the heavy quark mass is related to the perimeter law of the Wilson
loop and reminiscent of the classical Coulomb energy.
The instanton gas screens but does not confine.
Our construction is immediately amenable to heavy baryons.
The instanton size $\rho$ is fixed to $1/3$ fm throughout
and the instanton density
$n_*$ is fixed to 1 fm$^{-4}$.

{\bf 1.} Consider the following meson correlator in Minkowski space
\beq
{\cal C}_{\Gamma}(x,x') = \langle 0 | T \left (
\overline{q} \Gamma \psi (x) \overline{\psi} \Gamma q(x') \right ) |0 \rangle
\label{1}
\enq
with $\Gamma =({\bf 1}, \gamma )\times ({\bf 1}, T )$ an arbitrary
spin-flavor channel. In the regime where the mass $m_Q$ of the field $\psi$
is heavy, the correlator (\ref{1}) may be analysed systematically in
$1/m_Q$ by redefining the field $\psi$ using the Foldy-Wouthuysen (FW)
construction \cite{fw2}.
In the bare version and to leading order in $1/m_Q$,
the transformation reads
\beq
\psi (x) \sim e^{-i\gamma_0 m_Q t}\,\,
e^{-i\sigma_{0i}[\nabla^0, \nabla^i]/4m_Q^2}\,\,
e^{-i\vec\gamma\cdot\vec\nabla/2m_Q}\,\,Q (x)
\label{2}
\enq
where $\nabla =\partial - i A$ and with $g$ (the gauge coupling) set to one.
The first transformation  rescales the momenta, the second
eliminates the odd parts, and the third removes the mass term. The successive
transformations in (2) are vector-like, unitary and gauge-covariant. Thus,
they are
norm-preserving and anomaly-free. In terms of (\ref{2}) the QCD part of the
action for the heavy field  $\psi$ becomes
\beq
{\cal L}_{\psi} \sim \overline{Q}  i\gamma^0\nabla^0 Q  +
\overline{Q} \left(-\frac{\vec\nabla^2}{2m_Q} -\frac {\vec\sigma\cdot \vec B}
{2m_Q} \right) Q
\label{3}
\enq
with $B^i= -i\epsilon^{ijk} [\nabla^i, \nabla^j]$. Equation
(\ref{3}) has the expected
FW form to order $m_Q^{-1}$. Under (\ref{2}) the heavy meson source shifts
\beq
\overline{q}\Gamma \psi \sim \overline{q}\Gamma e^{-i\gamma^0 m_Q t}
\left( 1- \frac{i\vec\gamma\cdot\vec\nabla}{2m_Q}\right) Q
\label{4}
\enq
As a result, the correlator (\ref{1}) takes the generic form
\beq
{\cal C}_{\Gamma} \sim
\langle 0| \overline{q} \Gamma e^{-i\gamma^0 m_Q (t-t')} Q \overline{Q} \Gamma
q
|0 \rangle +
\langle 0| \overline{q} \Gamma e^{-i\gamma^0 m_Q (t+t')}
\bigg [ Q\overline{Q},
\frac{i\vec\gamma\cdot\stackrel{\leftarrow}{\nabla}}{2m_Q}\bigg ] \Gamma q
|0\rangle
\label{5}
\enq
This construction can be extended to arbitrary orders in $1/m_Q$.
Since the FW transformation preserves manifest gauge invariance order by order
in the heavy quark mass, the issue of renormalizability holds order by order in
$1/m_Q$. Our construction refers to the bare quark mass $m_Q$. We are assuming
that the expansion is not upset by renormalization.

We note that the second term in (\ref{5}) induces mixing between
the particle and the antiparticle content of the correlator. This
mixing follows from the redefinition of the sources and not
the action. As we are interested in the case where $m_Q$ is large (
non-relativistic limit ) it is convenient to take the projected
sources $\Gamma_{\pm} =\Gamma {(1\pm \gamma^0)}/2$ instead of the unprojected
sources $\Gamma$. Since $\Gamma_{\pm}
\vec\gamma\cdot\vec\nabla \Gamma_{\pm} =0$ the mixing part in (\ref{5}) drops.
Thus, in Euclidean space
\beq
{\cal C}^{\pm}_{\Gamma}(x,x')\sim - e^{\mp m_Q (\tau -\tau')}
\langle 0 | {\rm Tr} \left(\Gamma_{\pm} \,S_Q(x,x')\,\Gamma_{\pm} \, S(x' , x
)\right) |0 \rangle
\label{6}
\enq
where $S$ is the propagator of the light quark, and $S_Q$ is the
heavy quark propagator,
\beq
S_Q \sim S_{\infty} +S_{\infty}
\left(-\frac{\vec\nabla^2}{2m_Q} -\frac {\vec\sigma\cdot \vec B}{2m_Q}
\right) S_{\infty}  + O(m_Q^{-2})
\label{7}
\enq
with $S_{\infty}=\gamma_4/i\nabla_4$ being the free part.
The order $m_Q^0$ has been considered by many authors \cite{hqet}.
Terms of order $m^{-2}_Q$ and higher can be sought along the same lines.

The present construction has the advantage of transforming the complete
correlator and action into an effective correlator and an effective action
with manifest power counting in $1/m_Q$.
The transformation does not rely on the commonly used momentum decomposition
$p_Q = m_Q v + k$, and as such does not suffer from the ambiguities associated
with the redefinition of the heavy quark velocity $v$ \cite{luma}.
We have explicitly used
the rest frame of the heavy quark, thus $v= (1,\vec 0)$ in our case. The
derivation can be easily generalized to arbitrary frames.

{\bf 2.} We now proceed to evaluate (\ref{7}) to order $m_Q^0$ in a random gas
of instantons. In general, the correlator
${\cal C} \sim \langle S \otimes S_{\infty} \rangle$ receives contributions
from both planar and non-planar graphs and thus is difficult to analyze.
However, to leading order in $1/N_c$ the planar graphs
dominate \cite{diapob,pob}.
To this order, the $\eta^\prime$  is $massive$ (even in the
chiral limit) and the instantons are
screened. A dilute system of instantons and antiinstantons with zero net
topological charge behaves as a free gas. The system screens but does not
confine (see below). This notwithstanding, the planar graphs may be  resummed
using a Bethe-Salpeter-type equation. Following Pobylitsa \cite{pob},
the inverse correlator (\ref{6}) after resummation reads
\beq
{\cal C}^{-1} \sim S^{-1} \otimes S_{\infty}^{T\,-1}  - \sum_{I,\overline{I}}
\; \langle \,\,({S} -\Aslash_{I}^{-1}  )^{-1} \otimes
( {S}_{\infty} -\Aslash_{4,I}^{-1} )^{T\, -1}
 \,\,\rangle
\label{8}
\enq
The upper script $T$ is short for transpose.
In the planar approximation, the light propagator satisfies the
integral equation
\beq
S^{-1} = S_0^{-1} - \sum_{I,\overline I}
\langle\left( \,\,\Aslash_I^{-1} -S \right)^{-1}\rangle
\label{9}
\enq
For massless quarks $S_0^{-1} = +i\dslash$, while the infinitely heavy
quark propagator satisfies
\beq
S_{\infty}^{-1} = S_*^{-1} - \sum_{I, \overline I} \langle\left(
\,\,\Aslash_{4,I}^{-1} -S_{\infty} \right)^{-1}\rangle
\label{10}
\enq
where $\Aslash_4 =\gamma^4 A^4$ and $S_* =i\gamma^4\partial^4$.
The sum is over all instantons and antiinstantons and the averaging is
over their position $z_I$ and color orientation $U_I$. Generically
\beq
\sum_{I,\overline I} \rightarrow \frac N2 \,\,\left(\frac 1{V_4}\int
d^4z_I\right) \,\,\int dU_I  + \left( I\rightarrow \overline I \right)
\sim n\int d^4z \,\,\left( I +\overline I \right)
\label{11}
\enq
where $n=N/2V_4N_c=n_*/2N_c$.
The last substitution follows from color averaging over two $U_I$'s
for illustration. The diluteness factor of the
gas is given by the dimensionless combination $n\rho^4\sim 10^{-3}$, with
$n\sim N_c^0$ since $n_*\sim N_c$ \cite{dia}.

In the dilute gas approximation, the nested integral equations
(\ref{8}-\ref{10}) may be iterated in $n$. To leading order
\beq
{\cal C}^{-1} \sim
  S^{-1}  \otimes S_{\infty}^{T,\,-1} - n \int d^4 z \; {\rm Tr}_C
\left( [ \; L \; ]_I \; \otimes \; [ \; H \; ]_I + \,\,I\rightarrow{\overline
I}\right)
\label{12}
\enq
with
\beq
[ \; L \; ]_I\otimes [H]_I =\bigg [ S_0^{-1}\left(\,\,
\frac{| \Phi_0 \rangle \langle \Phi_0 |  }{i\sqrt{n}\Sigma_0} -S_0\,\,\right)
S_0^{-1} \bigg ]\otimes \bigg [ S_*^{-1} \left( \frac 1{i\gamma^4\nabla_{4I}}
-S_*\right)S_*^{-1}\bigg ]^T
\label{13}
\enq
Here $|\Phi_0\rangle$ is the normalized  fermionic
zero mode in the one-instanton
background, and
$\Sigma_0  = \langle \Phi_0 | \Sigma | \Phi_0 \rangle
\sim (240 \; {\rm MeV})^{-1}$.
We note that the effects of the planar resummation
is to generate an effective mass for the zero mode in (\ref{13}). This result
is directly related to the induced
constituent quark mass. Indeed, resumming the planar graph for the single quark
propagator shows that $S^{-1}\sim S^{-1}_0 +i\sqrt{n} \Sigma$, where $\Sigma$
is a space-spin-dependent kernel that satisfies the following
gap-like equation \cite{pob}
\beq
\Sigma = \int d^4z \,\,{\rm Tr}_C \left( S_0^{-1} \frac
{|\Phi_0><\Phi_0|}{\Sigma_0} S_0^{-1} +  I\rightarrow\overline I \right)
\label{14}
\enq
with $\Sigma_0$ following from (\ref{14}) by taking the expectation
value in the zero mode state. The shift in the light  quark mass is
\beq
\Delta M_q \sim \sqrt{n} <x_{-\infty} | \Sigma |x_{+\infty} > \sim
\sqrt{n} \left(\frac{4\pi^2\rho^2}{\Sigma_0}\right)
\sim 420 \,\,{\rm MeV}
\label{15}
\enq

Similarly for the heavy quark propagator, we have $S_{\infty}^{-1}\sim
S_*^{-1} + in \,\Theta$, where $\Theta$ is a
t-dependent kernel satisfying
\beq
\Theta = \int d^4z\,\,{\rm Tr}_C \left( S_*^{-1}
\left( \frac 1{i\gamma^4\nabla_{4,I}}-S_*\right) S_*^{-1} +
 I\rightarrow\overline I \right)
\label{16}
\enq
In coordinate space, the heavy quark propagator in the one instanton background
reads
\beq
<x|\frac 1{i\nabla_{4,I}} |0> =\delta (\vec x )\,\,\theta (\tau )\,\,
{\bf P}e^{i\int_0^{\tau}\, dsA_4(x_s-z_I )}
\label{17}
\enq
with $x_s = (s,\vec x )$.
Inserting (\ref{17}) into (\ref{16}) and using the one-instanton configuration
\beq
A_{\mu}^a (x) = +{\overline\eta}^a_{\mu\nu} x_{\nu}
\left( \frac 1{x^2} - \frac 1{x^2 +\rho^2 }\right)
\label{18}
\enq
yields for large times
\beq
<x_{-\infty} | \Theta | x_{+\infty} > \sim
\left( -32\pi \rho^3 \gamma^4\,\right) \int_0^{\infty} \left(x\,
{\rm cos}(\frac{\pi x}{2\sqrt{1+x^2}})\right)^2  =
\left( -32\pi\rho^3  \gamma^4 \,I_Q \right)
\label{19}
\enq
The corresponding shift  in the heavy quark mass follows from the effective
vertex (16)
\beq
\Delta M_Q = 32\pi n \rho^3 I_Q \sim 16\pi n\rho^3 \sim 70 \,\,{\rm MeV}
\label{20}
\enq
as first suggested by Diakonov, Petrov and Pobylitsa \cite{diapob}.
Note that the shift in the  heavy quark mass is of order $n$
as opposed to the shift in the light quark mass which is of order $\sqrt{n}$,
almost an order of magnitude smaller. This is simply the
classical Coulomb energy of a  heavy quark. A similar result has been recently
obtained by Bigi, Shifman, Uraltsev and Vainshtein and also
S. Narison  \cite{bigi}.
Remarkably, their estimate
using the infrared renormalon suggests $\Delta M_Q \sim 50$ MeV.
and $\Delta M_Q \sim 70$ MeV respectively.

{\bf 3.} The shift in the heavy quark mass to order $m_Q^0$
is related to the behavior of large Wilson loops in the random instanton gas.
Indeed, the Wilson loop can be rewritten as follows
\begin{equation}
\langle {\rm Tr}_c{\bf P}e^{i\int_{T\times L} dx\cdot A} \rangle = -
\bigg < {\rm Tr}_c\left( {\bf P}e^{i\int_{T_1} dx\cdot A (T_1)}\,
{\bf P}e^{i\int_{L_1} dx\cdot A (L_1)}\,{\bf P}e^{i\int_{T_2} dx\cdot A
(T_2)}\,
{\bf P}e^{i\int_{L_2} dx\cdot A (L_2)} \right)\bigg >
\label{w1}
\end{equation}
where $T_{1,2}$ and $L_{1,2}$ are the paths shown in Fig. 1.
$A (T,L) =\sum_I A_I (T, L)$ is the sum of the instanton and antiinstanton
gauge configurations projected onto the lines $T,L$. For large Wilson loops
$T, L >> 1/\rho$, the heavy quarks (line integrals) $no$ $longer$ interact
with each other. As a result, (\ref{w1}) factorizes into the trace of the
product of four "heavy quark" propagators (Wilson lines), each averaged
over the ensemble of
instantons and antiinstantons, and evaluated for large separations
$T$ and $L$ respectively. The above discussion shows that the heavy quark
propagator acquires a mass $\Delta M_Q$ asymptotically. Thus
\beq
<{\rm Tr}_c{\bf P}e^{i\int_{T\times L} dx\cdot A} > \sim -
e^{-\Delta M_Q (2T+2L)}
\label{w2}
\enq
Large Wilson loops obey a perimeter law with $\sigma_P =\Delta M_Q$.
By restricting the discussion to classical fields the nasty issue
of the divergences in the Coulomb energy does not arise. It would be
interesting to find out, how (\ref{w2}) compares with unquenched
lattice calculations  before and after cooling.
A similar behavior is also present in the Schwinger model.

In QCD, large Wilson loops are  expected to obey a perimeter law. Does this
mean that dilute instanton systems reflect faithfully  on unquenched QCD ?
We do not think so. In the presence of light quarks, the instantons and
antiinstantons in the vacuum are screened over distance scales on the order of
 $1/2$ fm.
The screening is due to feedback of the quarks on the instanton
configurations. While the effect is naively down by $1/N_c$, since the
instanton density grows with $N_c$, the screening persists to order $N_c^0$.
This screening causes the topological
susceptibility to be small and the $\eta^\prime$ to be massive.
Dilute instanton systems in the presence of light quarks
behave as a free gas.
The operating mechanism is $screening$ and not $confinement$.
The lack of confinement is dramatically seen in (\ref{8}) where the first
term in the inverse correlator reflects on a dressed heavy-light quark cut.
This problem occurs in all correlators.

{\bf 4.} The above notwithstanding, the inverse correlator (\ref{13})
allows for an immediate
interpretation in terms of effective interactions. In the long wavelength
limit, the instanton size is small, and a local effective interaction between
the effective fields ${\bf Q}$ and ${\bf q}$ (as opposed to the "fundamental"
fields $Q$ and $q$) can be derived much like
the 't Hooft interaction between the light effective fields ${\bf q}$.
To show this, let $(x, \; y)$ and $(x^\prime , \; y^\prime )$
be the coordinates of the light and heavy quarks respectively before and after
they have interacted with an instanton (antiinstanton).
{}From Bethe-Salpeter equation we read the vertex
\beq
\Gamma( x,y, \; x^\prime , y^\prime ) =
- i n N_c \int d^4 z \; \int dU \left (
\langle x | [ \; L \; ]| x^\prime \rangle
\otimes \langle y | [\; H \; ] | y^\prime \rangle
+ I \rar \overline{I} \right ) \label{22}
\enq
This vertex function gives rise to an effective action ${\cal S}_I$.
Averaging over the instanton orientation and position yields in the
long wavelength (local) approximation
\beq
{\cal L}_{qQ} = n \bigg (- \frac {16\pi\rho^3 I_Q}{N_c} \bigg )
\bigg (\frac{4\pi^2\rho^2}{\sqrt{n}\Sigma_0} \bigg ) \left (
i{\bf Q}^\dagger \frac{1+i\gamma_4}2 {\bf Q}\,\,i{\bf q}^\dagger {\bf q} +
\frac{1}{4} i{\bf Q}^\dagger \frac{1+i\gamma_4}2 \lambda^a {\bf Q}\,\,
i{\bf q}^\dagger  \lambda^a {\bf q} \right )
\label{24}
\enq
The first bracket in (\ref{24}) arises from the heavy quark part and the second
bracket from the light quark part. Wick-rotating to Minkowski space gives
\beq
{\cal L}_{qQ} = -\bigg (\frac {\Delta M_Q\Delta M_q}{2nN_c} \bigg )\,\,
\left ( \overline{{\bf Q}} \frac{1+\gamma^0}2 {\bf Q}\,\,
\overline{{\bf q}} {\bf q} +
\frac{1}{4} \overline{{\bf Q}} \frac{1+\gamma^0}2 \lambda^a {\bf Q}\,\,
\overline{{\bf q}} \lambda^a {\bf q} \right )
\label{25}
\enq
which is to be compared with the 't Hooft vertex for two light
flavors ${\bf q}=({\bf u}, {\bf d})$
\beq
{\cal L}_{qq} = \bigg ( \frac {\Delta M^2_q}{nN_c} \bigg ) \,\,
\left( {\rm det }\,\overline {\bf q}_R {\bf q}_L \,\,+\,\,{\rm det}\,
\overline {\bf q}_L {\bf q}_R\right)
\label{26}
\enq
The ratio of the strengths is
${\kappa_{qQ}}/{\kappa_{qq}} = {\Delta M_Q}/2{\Delta M_q}\sim 0.08$ and
of order $N_c^0$, with $\kappa_{qQ}\sim \sqrt{n} /N_c$.

Similar
arguments may be applied to $\overline{Q} Q$ mesons as well.
The induced effective Lagrangian for heavy-light and
heavy-heavy mesons is given by
\beq
{\cal L} &=& \overline {\bf q} \left( i\dslash - \Delta M_q \right) {\bf q} +
           \overline {\bf Q} \frac {1+\gamma^0}2
             \left( i\partial_t -\Delta M_Q\right) {\bf Q}\nonumber\\
&-& \kappa_{qQ} \,\,\left(\overline {\bf Q} \frac{1+\gamma^0}2 {\bf Q}
               \,\,\overline {\bf q} {\bf q}
 + \frac 14 \overline {\bf Q} \frac{1+\gamma^0}2 \lambda^a {\bf Q}
               \,\,\overline {\bf q} \lambda^a {\bf q}\right) \non \\
&-& \kappa_{QQ} \,\,\left(\overline {\bf Q} \frac{1+\gamma^0}2 {\bf Q}
               \,\,\overline {\bf Q} \frac{1+\gamma^0}2 {\bf Q}
 +  \frac 14 \overline {\bf Q} \frac{1+\gamma^0}2 \lambda^a {\bf Q}\,\,
                \overline {\bf Q} \frac{1+\gamma^0}2 \lambda^a {\bf Q}\right)
\label{28}
\enq
where $\kappa_{QQ} =\Delta M_Q^2/2nN_c $.
For heavy baryons the construction applies as well. For $qqQ$ systems the
answer in Minkowski space reads
\beq
{\cal L}_{qqQ} = -\bigg ( \frac{\Delta M_Q \Delta M_q^2}{2n^2 N^2_c} \bigg )
\bigg ( &&\overline{\bf Q} \frac {1+\gamma^0}2 {\bf Q} \,\,
\left({\rm det} \overline {\bf q}_L {\bf q}_R \,\,+{\rm det}\overline
{\bf q}_R {\bf q}_L \,\,\right) +\nonumber\\ &&
       \frac 14 \,\,\overline{\bf Q} \frac {1+\gamma^0}2 \lambda^a {\bf Q} \,\,
\left({\rm det} \overline {\bf q}_L\lambda^a {\bf q}_R \,\,+{\rm det}\overline
{\bf q}_R\lambda^a {\bf q}_L \,\,\right) \bigg )
\label{29}
\enq
We note that $\kappa_{qqQ}\sim n^0 /N^2_c$.
A comparison with the conventional 't Hooft determinantal interaction
shows that $\kappa_{qqQ}/\kappa_{qqq} = \kappa_{qQ}/\kappa_{qq}\sim 0.08$.
The ratio of the strengths in heavy to light baryons is the same as the ratio
of the strengths in heavy to light mesons.
We note that both (\ref{28}) and (\ref{29}) are invariant under a spin
flip of the heavy quark $SU(2)_Q$ and $U_A(1)$ violating, and that
(\ref{29}) is chiral $SU(2)_L\times SU(2)_R$ invariant.

In terms of the effective quark fields ${\bf Q}$ and ${\bf q}$ the dynamics is
the one of the constituent quark model to order $m_Q^0$. For heavy light
mesons, the spectrum is
\beq
M_{qQ} = (m_Q +\Delta M_Q) + (m_q + \Delta M_q) +\alpha_{qQ}\,\,\kappa_{qQ}
\left({\bf 1} + \frac 14 \lambda_q^a \cdot \lambda_Q^a \right)
\label{30}
\enq
with $\alpha_{qQ} = |\phi_{qQ} (0)|^2
\sim 1/a_{qQ}^3$, where $\phi_{qQ} (0)$ is the $qQ$-wavefunction at the origin
and $a_{qQ}$  its typical size. Using the Van-Royen-Weisskopf construction
\cite{weis}, we obtain
\beq
a_{qQ}^2 \sim \frac{N_c}{2 \pi^2 f_{qQ}^2}
\label{31}
\enq
with $a_{qQ} \sim 0.26$ fm for $f_D \leq 290$MeV \cite{hqet}.
The spectrum (\ref{30}) follows readily from the large distance
asymptotics of the heavy-light  correlator
(\ref{12}) if we were to use the induced vertices (\ref{25}) for simplicity.
The interaction part in (\ref{30}) corresponds to the scalar  (${\bf 1}$)
and the (instanton-induced) Coulomb-like-exchange ($\lambda\cdot\lambda$).
The latter is leading and attractive in singlet configurations. Indeed,
since the Casimir in the fundamental representation
($C_F =\lambda\cdot\lambda/4$) scales with $N_c$, the shift in the heavy-light
 mass
is of order $N_c^0$ with $\alpha_{qQ}\sim N_c^0$ and $\kappa_{qQ}\sim 1/N_c$,
\beq
\Delta M_{qQ} \sim -{C_F}\,\,{\alpha_{qQ} \kappa_{qQ}}
              \sim -\frac {N_c}2\,\,{\alpha_{qQ} \kappa_{qQ}}
\label{32}
\enq
Recall that (\ref{26}) in the large $N_c$ limit  gives the following binding
energy
\beq
\Delta M_{qq}\sim   -(1-4\,S_q\cdot S_q) \,{C_F}
                      \alpha_{qq} \kappa_{qq}
\label{33}
\enq
For spin zero mesons $1-4S_q\cdot S_q = 4$ so that
$\Delta M_{qq}\sim -2N_c\alpha_{qq}\kappa_{qq}$. With this in mind, and
using (\ref{31}), the ratio of the bindings in heavy-light to
light-light systems reduces to
\beq
\frac {\Delta M_{qQ}}{\Delta M_{qq}} \sim \frac 18\bigg (
\frac{\Delta M_Q}{\Delta M_q}\bigg )
\bigg ( \frac{f_{qQ}}{f_{qq}} \bigg )^3
\label{34}
\enq
and similarly for the ratio of the bindings for heavy-heavy to light-light
systems
\beq
\frac {\Delta M_{QQ}}{\Delta M_{qq}} \sim \frac 18\bigg (
\frac{\Delta M_Q}{\Delta M_q}\bigg )^2
\bigg ( \frac{f_{QQ}}{f_{qq}} \bigg )^3
\label{35}
\enq
For D-mesons, we have $\Delta M_D \sim 0.22\,\, \Delta M_{\pi}$
($f_D \sim 290$ MeV), while for
charmonium $\Delta M_J \sim 0.10\,\,\Delta M_{\pi}$ ($f_J \sim 410$ MeV).
The drop in
the interaction strength for heavy-heavy systems is compensated by an
increase in the decay constant, in comparison to light-light systems.
Since the light constituent mass is about 420 MeV,
the pion binding energy is about 700 MeV. Thus
the binding energy in heavy-light and heavy-heavy systems is about a
100 MeV. These estimates are consistent with the Coulomb estimates in the
non-relativistic quark model \cite{deruj}.

Similar relations and estimates
hold for baryons. They will be discussed elsewhere along with hyperfine
splittings. Subleading effects in $1/N_c$ arise from the left out parts
in the color averaging, the non-planar contributions and the perturbative
gluon exchanges. They are usually harder to assess without further assumptions.

{\bf 5.} We have explicitly shown how the heavy meson correlator can be
analyzed in powers of $1/m_Q$ prior to radiative corrections. Using a random
and dilute instanton gas, we have shown that the heavy-light correlator can
be calculated in the planar approximation and to leading order in the density
fairly accurately. In the random gas the heavy and light quarks acquire a mass
to order $m_Q^0$. The induced mass of the heavy quark is related to the
perimeter law displayed by the large Wilson loops. The latter reflects on
screening in the gas. We have shown that in the long wavelength limit,
the interaction induced by instantons between heavy and light quarks
is attractive, and both chiral and heavy-quark symmetric. Our construction
is immediately amenable to heavy baryons. It also allows for  systematic
investigations of the $1/m_Q$ corrections and a possible assessment of the
$1/N_c$ corrections. Some of these issues will be taken up in the future.

\vskip 1cm

{\bf Acknowledgement}

This work has been supported in part by a DOE Grant No.\, DE-FG02-88ER40388.

\vfill
\newpage

\setlength{\baselineskip}{18pt}

\end{document}